\documentclass[5p]{elsarticle}

\usepackage[american]{babel}
\usepackage[separate-uncertainty=true]{siunitx}
\usepackage{xcolor}
\usepackage{graphicx}
\usepackage{float}
\usepackage{amsmath}
\usepackage{amssymb}
\usepackage[normalem]{ulem}
\usepackage[modulo, mathlines]{lineno}

\let\oldequation\equation
\let\oldendequation\endequation

\renewenvironment{equation}
  {\linenomathNonumbers\oldequation}
  {\oldendequation\endlinenomath}

\usepackage[breaklinks=true, colorlinks]{hyperref}

\newcommand{\Nix}[2]{\ensuremath{\text{(CoCrFeMn)}_{#1}\allowbreak\text{Ni}_{#2}}}

\begin{document}

\title{Thermodynamics of vacancies in concentrated solid solutions:\\
From dilute Ni-alloys to the Cantor system}

\author[mm]{Daniel Utt \corref{cor1}}
\ead{utt@mm.tu-darmstadt.de}
\cortext[cor1]{Corresponding author:}
\author[mm]{Alexander Stukowski}
\author[mm]{Karsten Albe}
\address[mm]{Technische Universit\"at Darmstadt, Otto-Berndt-Str.\ 3, 64287 Darmstadt, Germany}

% \date{\today}

\begin{abstract}
The equilibrium concentration of vacancies in various \Nix{1-x}{x} alloys is studied at finite temperature by grand-canonical lattice Monte Carlo (MC) simulations. The formation energies are calculated from a classical interatomic potential and exhibit a distribution due to the different chemical environments of the vacated sites. In dilute alloys, this distribution features multiple discrete peaks, while concentrated alloys exhibit an unimodal distribution as there are many different chemical environments of similar vacancy formation energy. MC simulations using a numerically efficient bond counting model confirm that the vacancy concentration even in concentrated alloys may be calculated by the established Maxwell-Boltzmann equation weighted by the given distribution of formation energies. We calculate the variation of vacancy concentration as function of Ni content in the \Nix{1-x}{x} alloy system and prove the excellent agreement of the thermodynamic model and the results from the grand-canonical Monte Carlo simulations. We further show that the apparent vacancy formation energy obtained from an Arrhenius-type presentation of the temperature dependent concentrations differs massively from the mean value of distribution of the formation energies.
\end{abstract}

\begin{keyword}
  high-entropy alloy \sep vacancy thermodynamics \sep vacancy concentration
\end{keyword}

\maketitle

\section{Introduction}
Vacancies in densely packed metals and alloys govern a wide range of materials properties, including electrical and thermal conductivity, ordering, segregation and creep phenomena \cite{Dienes1951,Seah1980,Fiala1991}. While an in-depth understanding of vacancy thermodynamics and kinetics exists for dilute \cite{Porter2009} and binary alloys \cite{Zhang2015a, Ruban2016, Morgan2020}, there is no consensus yet how equilibrium vacancy concentrations can be theoretically predicted for concentrated solid solutions including high-entropy alloys (HEAs).
There are two main culprits making the calculation of the vacancy concentration in non-dilute alloys difficult: Firstly, the vacancy formation energy is not a single value but an energy distribution since every vacant site has a different chemical environment \cite{Ruban2016}. In CoCrFeMnNi, for example, vacancy formation energies can scatter between \SI{1.6}{\electronvolt} and \SI{2.4}{\electronvolt} \cite{Mizuno2019} or \SI{1.51}{\electronvolt} and \SI{2.72}{\electronvolt} \cite{Guan2020}.
Secondly, it is still unclear, how the vacancy concentration can be determined from these vacancy formation energies. Here, the proper treatment of the configurational entropy and its impact on the equilibrium vacancy concentration is the main point of contention.
While Wang et al.\ \cite{Wang2017} state that the configurational entropy inherent to HEAs leads to an increase in equilibrium vacancy concentration compared to monoatomic metals with the same vacancy formation energy, Ruban~\cite{Ruban2016} argue that in an equimolar alloy of \(\mathcal{N}\) constituents the vacancy concentration is reduced by a factor of \(1/\mathcal{N}\). Lastly, Morgan and Zhang~\cite{Morgan2020} oppose both statements and report that the equilibrium vacancy concentration is independent of the number of constituents in an alloy.

These diverging views call for a closer inspection of the equilibrium thermodynamics of vacancies in multi-component alloys. Hence, in this study we present atomistic calculations on the concentration dependent vacancy formation energies in the face-centered cubic (FCC) \Nix{1-x}{x} HEA system and determine the resulting vacancy concentration using grand-canonical lattice Monte Carlo simulations. The goal is to assess to what extent the composition of the host alloy influences the equilibrium vacancy concentration and to determine which of the three aforementioned thermodynamic models provides a correct description of the vacancy concentration in multi-component alloys. Furthermore, we assess the impact of the energy distribution on the ``apparent''  vacancy formation energy as determined from an Arrhenius representation.

\section{Methodology}
The vacancy formation energies of the \Nix{1-x}{x} series of alloys were obtained from a second nearest-neighbor modified embedded-atom method (MEAM) interatomic potential \cite{Lee2000} parametrized by Choi et al.\ \cite{Choi2018}. The static energies were calculated via energy minimization algorithms implemented in \textsc{lammps} \cite{Plimpton1993}. The ideal random solid solution samples were prepared using \textsc{atomsk} \cite{Hirel2015}, visualization and further processing was performed in \textsc{ovito} \cite{Stukowski2010}. 

\subsection{Calculation of vacancy formation energies from molecular statics}

The Gibbs free energy of vacancy formation for a specific lattice atom \(i\) in a multi-component alloy with \(N\) sites is given by
\begin{equation}
  G_\text{f} = H_\text{f} - TS_\text{f} = G_\text{Def,i}^{(N-1)}  - \left( G_\text{Ref}^{(N)} - \mu_i \right),\label{eq_diff_vacEngG}
\end{equation}
where \(G_{\text{Def},i}^{(N-1)}\) is the free energy of the simulation cell containing the vacancy and \(G_\text{Ref}^{(N)}\) is the energy of the system with \(N\) occupied sites without the vacancy.  The chemical potential of the reservoir for the removed species \(i\) is \textbf{\(\mu_i\)}. 
If we neglect the formation entropy \(S_{\text{f}}\) and the pressure dependence, the formation energies can be approximated by the  potential energy \(E\). Therefore, \autoref{eq_diff_vacEngG} may be rewritten as
\begin{equation}
  G_\text{f} \approx E_\text{f} = E^{(N-1)}_{\text{Def},i}  - \left( E^{(N)}_\text{Ref} - \mu_i \left[ T=\SI{0}{\kelvin} \right] \right) .\label{eq_diff_vacEng}
\end{equation}

The chemical potential \(\mu_i\)  corresponds to the change in Gibbs free energy \(G\) resulting from a change in particle count of species \(i\)
\begin{equation}
  \mu_i = \left( \frac{\partial G}{\partial N_i} \right)_{T,P,N_{i}}%
  \xrightarrow[]{T = \SI{0}{\kelvin}}%
  \left( \frac{\partial E}{\partial N_i} \right)_{T,P,N_{i}},\label{eq_diff_chemPot}
\end{equation}
where \(N_i\) denotes the number of particles of type \(i\) and \(E\) is the potential energy calculated by \textsc{lammps}. 
We obtained \(\mu_i\) at \(T=\SI{0}{\kelvin}\) from a quasi-binary approach. In order to determine \(\mu_\text{Co}\), for example, \mbox{Co\(_{N_\text{Co} + \Delta N_\text{Co}}\)Cr\(_{N_\text{Cr}}\)Fe\(_{N_\text{Fe}}\)Ni\(_{N_\text{Ni}}\) } samples with a varying number of Co atoms were created. Here, \(N_i\) is the number of atoms per species \(i\) and \(\Delta N_i\) is a small change in atom counts. Following energy minimization, the slope of the energy change \(\Delta E\) over \(\Delta N_i\) gives the chemical potential, assuming that the entropy contributions stay constant in this small window of compositional change.

An important fact to note at this point is that the thermodynamically relevant distribution of formation energies must not depend on the removed atom, since all species can occupy all possible lattice sites and thus we cannot distinguish different vacancy types. Even though, the terms \(E^{(N-1)}_{\text{Def},i}\) and \(\mu_i\) depend on the removed atom type \(i\) the distribution of values for \(G_\text{f}\) does not, as will also be shown later in the results part (\autoref{fig1} \& \autoref{fig2}). This is obvious, as the chemical potential of a vacancy in the lattice has to be independent of its lattice site and the removed atom type.

To calculate the vacancy formation energies simulation cells containing \num{e3} FCC unit cells for a total of \num{4000} lattice sites were built. For each \Nix{1-x}{x} sample \num{40} random atomic configurations were created. In each of these \num{40} samples, \num{40} separate vacancy sites of each of the 5 species were sampled resulting in \num{8000} vacancy calculations per alloy. This was done for a total of \(\approx \num{90000}\) individual calculations.
All energies were obtained under static conditions using conjugate gradient (CG) and \textsc{fire} \cite{Guenole2020} energy minimization. The simulation cell volume was kept constant at the defect free equilibrium volume and periodic boundary conditions were applied in all dimensions \cite{Mishin2001}. 

The samples used for determining the chemical potentials contained a varying number of lattice sites equal to \(\num{32000} \pm \Delta N_i\) atoms, with \(\Delta N_i \leq 192\). Note, that these samples were defect free with a varying number of lattice sites. The energy was minimized using the CG algorithm allowing for an anisotropic change in box size to reach pressure-free conditions.

\subsection{Vacancy concentrations from Monte-Carlo simulations using a bond-counting model\label{sec_vac_meth_conc}}
Due to the differences in chemical environments in concentrated random alloys, the Gibbs free energy of vacancy formation \(G_\text{f}\) is not a single value but instead follows a distribution \(g\left( G_\text{f} \right)\) \cite{Ruban2016}. 
Therefore, the vacancy concentration has to be obtained from integration of this energy distribution 
\begin{equation}
  c_\text{Vac} = \int g \left( G_\text{f} \right) \exp \left( - \frac{{G}_\text{f}}{k_\text{B} T} \right) d {G}_\text{f},\label{eq_cvac_gG}
\end{equation}
where \(k_\text{B}\) is the Boltzmann constant. 
According to Morgan and Zhang~\cite{Morgan2020}, this expression can be approximated by
\begin{equation}
    c_\text{Vac} = \exp \left( \frac{S_\text{f}}{k_\text{B}} \right) \int g \left( E_\text{f} \right) \exp \left( - \frac{E_\text{f}}{k_\text{B}T} \right) d E_\text{f}\label{eqVacM},
\end{equation}
if one assumes that the formation entropy is site independent.

There are, however, other views on the theoretical treatment of the vacancy concentration reported in literature. Ruban~\cite{Ruban2016}, for example, proposed to include the configurational entropy of the defect free host lattice. For an equimolar alloys, this leads to a factor \(1/\mathcal{N}\) in front of \autoref{eqVacM}, implying that the vacancy concentration should scale with the number of components \(\mathcal{N}\).  Wang et al.~\cite{Wang2017}, in contrast, argue that next to the number of components only the mean formation energy is a necessary descriptor and suggest yet another analytic form.

In light of this situation, we decided to apply an independent numerical scheme  and used grand-canonical (GC) Monte-Carlo (MC) simulations to calculate the temperature dependent vacancy concentrations. The simulations were run based on a custom build \textsc{julia} \cite{Bezanson2017} code which takes the site dependent vacancy formation energies from a bond-counting (BC) model to become computationally efficient.

The approach is based on the Metropolis algorithm, where in each MC trial step a lattice site is randomly selected and its vacancy formation energy is calculated. The atom removal, i.e., vacancy insertion, is accepted with a probability, \(P = \exp \left(-E_\text{f} / \left(k_\mathrm{B} T \right) \right)\). If a vacant site is selected, the original atom is always reinserted into the lattice to preserve the global sample composition. This is necessary, as the BC model specifically fit for each alloy composition and therefore not transferable across \(x_\mathrm{Ni}\).

The MC simulations were carried out with \num{24} independent samples for each temperature containing \num{4000} lattice sites. At this relatively small lattice size no more than one vacancy is present in the sample and  thus we avoid the formation of divacancies as they are not captured by the bond counting model. At \SI{1000}{\kelvin} each of the samples was simulated for \num{8e8} MC trial steps, while the higher temperature (\SIrange{1200}{1600}{\kelvin}) samples were simulated for \num{4e8} MC steps, each. We also simulated an \SI{800}{\kelvin} sample, however, convergence proved unattainable. For reference, \autoref{fig4}(a) shows the cumulative average of the vacancy concentration over MC steps for {different temperatures in the equimolar (\Nix{1-x}{x}) alloy. Good convergence of the MC simulation can be seen for temperatures of \SI{1000}{\kelvin} and above}.

\subsection{Bond counting model}

Since on-the-fly analysis of formation energies using the MEAM potential is computationally too demanding, we decided to fit a linear bond counting (BC) model to the reference data for describing the vacancy formation energy in a given chemical environment. This model uses the first and second nearest neighbor bonds around a vacancy as descriptors. The number of nearest (\(N_{i\text{-}j,1}\)) and second nearest (\(N_{i\text{-}j,2}\)) \(i\text{-}j\) neighbor bonds between particles of type \(i\) and \(j\) are counted. An example of such a chemical environment is shown in \autoref{fig3}(b). The environment around a vacancy is an adequate descriptor of the vacancy formation energy as it is independent of the removed atom's type (see \autoref{fig2} for details). 

The fitting is done using the Moore-Penrose pseudo-inverse method \cite{Strang1988} as implemented in \textsc{numpy} \cite{VanDerWalt2011,Oliphant2015}. The formalism is encapsulated in the following equation

{\tiny
\begin{align*}
\begin{bmatrix}
  N^1_{\text{Co-Co},1} & N^1_{\text{Co-Co},2} & N^1_{\text{Co-Cr},1} & \ldots & N^1_{\text{Ni-Ni},2} \\
  N^2_{\text{Co-Co},1} & \ddots & & & \vdots \\
  \vdots & & & \ddots & \vdots \\
  N^n_{\text{Co-Co},1} & & \ldots & & N^n_{\text{Ni-Ni},2}
\end{bmatrix}
\begin{bmatrix}
  \epsilon_{\text{Co-Co},1} \\
  \epsilon_{\text{Co-Co},2} \\
  \epsilon_{\text{Co-Cr},1} \\
  \vdots \\
  \epsilon_{\text{Ni-Ni},2}
\end{bmatrix}
&=
\begin{bmatrix}
  E^1_\text{f} \\[0.75em]
  E^2_\text{f} \\
  \vdots \\
  E^n_\text{f}
\end{bmatrix}
,
\end{align*}
}
indicating that \(n\) different vacancies are subsequently used to fit the interaction parameters \(\epsilon_{i\text{-}j,1}\) and \(\epsilon_{i\text{-} j,2}\) for a given composition.

For each \Nix{1-x}{x} composition, we had already calculated \(8000\) different vacancy formation energies (cf.\ \autoref{fig1}). This data was split into \SI{80}{\percent} `training set', used to fit \(\epsilon_{i\text{-}j}\), and \SI{20}{\percent} `test set' to validate the fit. The resulting best fit is then used in the MC calculations. 

\section{Results}
\subsection{Chemical potentials}

\begin{figure*}[tbp]
\centering
\includegraphics{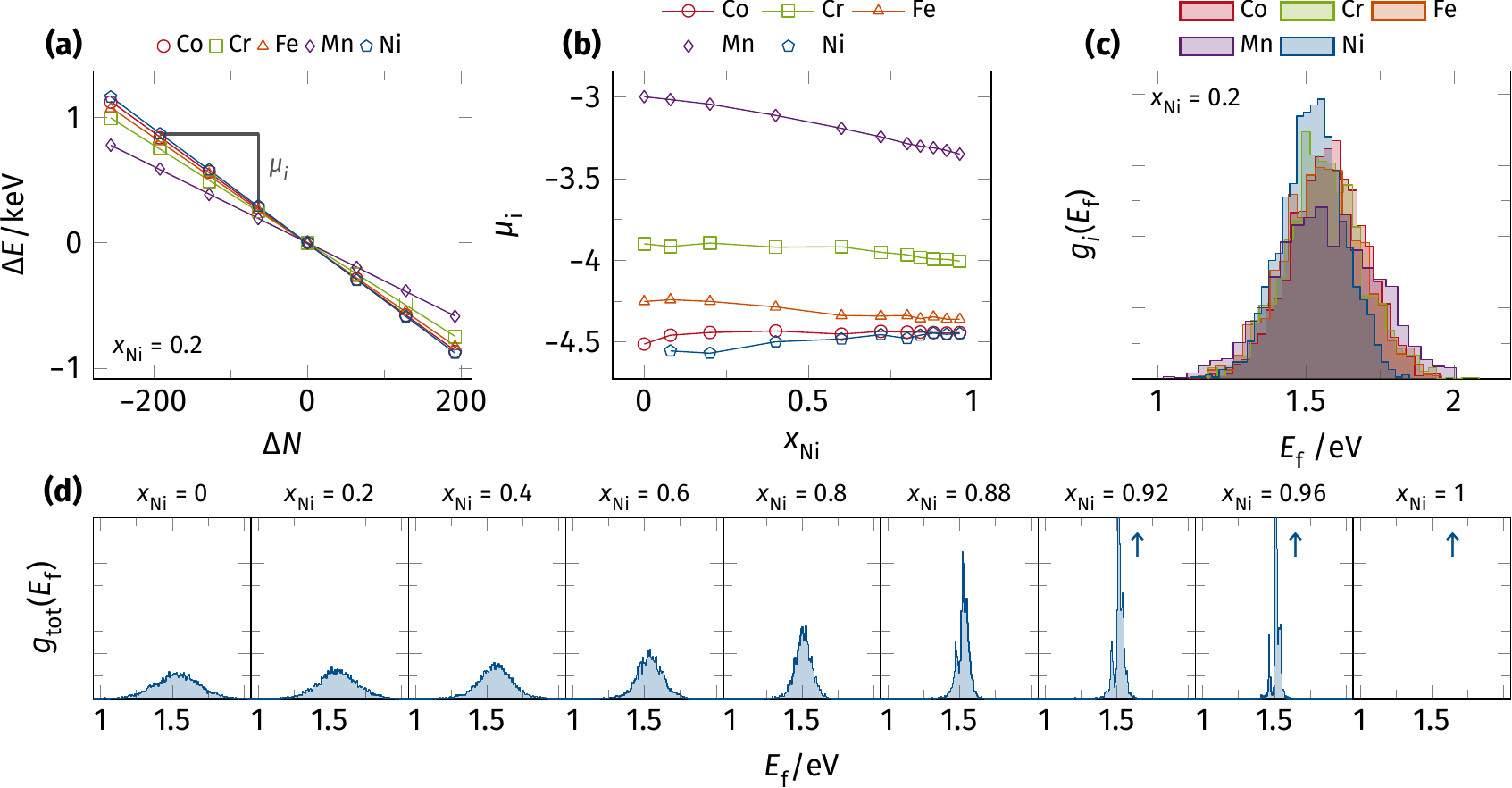}
\caption{
  (a) Chemical potential \(\mu_i\) calculated for the equimolar CoCrFeMnNi sample from the slope of the energy variation over particle number of species \(i\). 
  (b) Dependence of \(\mu_i\) on the Ni concentration \(x_\text{Ni}\) in the \Nix{1-x}{x} samples.
  (c) Distribution \(g_i\) of the vacancy formation energies \(E_\mathrm{f}\) resolved by the species {\(i\)} of the removed atom using chemical potentials from (b). The vacancy formation energy only depends on the chemical environment and not the species which was removed.
  (d) Distribution of total vacancy formation energies \(g_\text{tot} = \sum_i^N x_i g_i\), for the different Ni concentrations showing a transition from one broad peak for the concentrated alloys to multiple sharp peaks for the dilute Ni alloy. An \(\uparrow\) indicates that the peak of the distribution extends to higher \(g\)-values.
\label{fig1}}
\vspace{4em}
\includegraphics{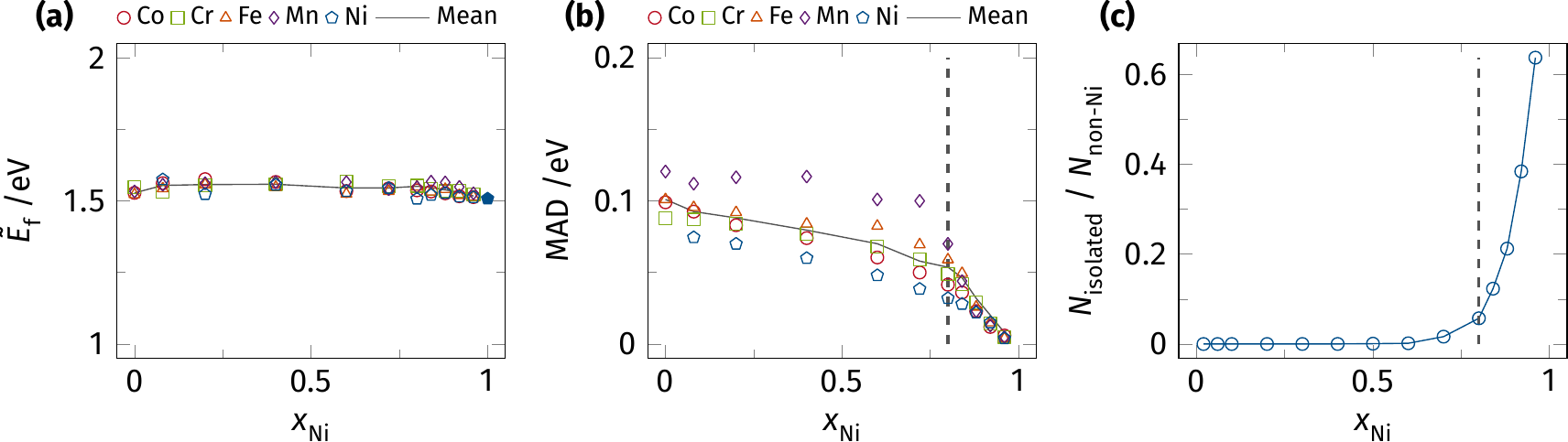}
\caption{
  (a) Median vacancy formation energy \(\widetilde{E}_\text{f}\) as function of Ni concentration \(x_\mathrm{Ni}\) in the different pseudo-binary \Nix{1-x}{x} alloys.
  (b) Median average deviation (MAD) \cite{Leys2013} taken as measure of the width of the respective vacancy formation energy distribution.
  Both \(\widetilde{E}_\text{f}\) and MAD are extracted from \autoref{fig1} (d). A dashed line indicates the percolation threshold on the FCC lattice marking the transition from dilute to concentrated solid solution.
  (c) Ratio of isolated non-Ni atoms \(N_\text{isolated}\) to the total number of non-Ni atoms \(N_\text{non-Ni}\) in different \Nix{1-x}{x} HEAs. The theoretical FCC percolation at \(x_\mathrm{Ni} \approx 0.8\) threshold\cite{Gaunt1983} is indicated.
  \label{fig2}}
\end{figure*}

In a first step, the chemical potential of each component \(i\) in the host material \(\mu_i\) is determined for all compositions of the \Nix{1-x}{x} system. This quantity is required to calculate the vacancy formation energy \(E_\text{f}\). \autoref{fig1}(a) shows the energy change with the variation of the particle number \(\Delta N_i\) for all species \(i\) in the equimolar CoCrFeMnNi HEA (cf.\ \autoref{eq_diff_chemPot}). The slope of the linear fit is used to determine the chemical potential. The resulting dependence of \(\mu_i\) on the Ni concentration \(x_\mathrm{Ni}\) is shown in \autoref{fig1}(b), which indicates that the chemical potential of Mn is most sensitive to the Ni content within this family of alloys, while the other components show fairly constant chemical potentials.

\begin{figure*}[tb]
\centering
\includegraphics{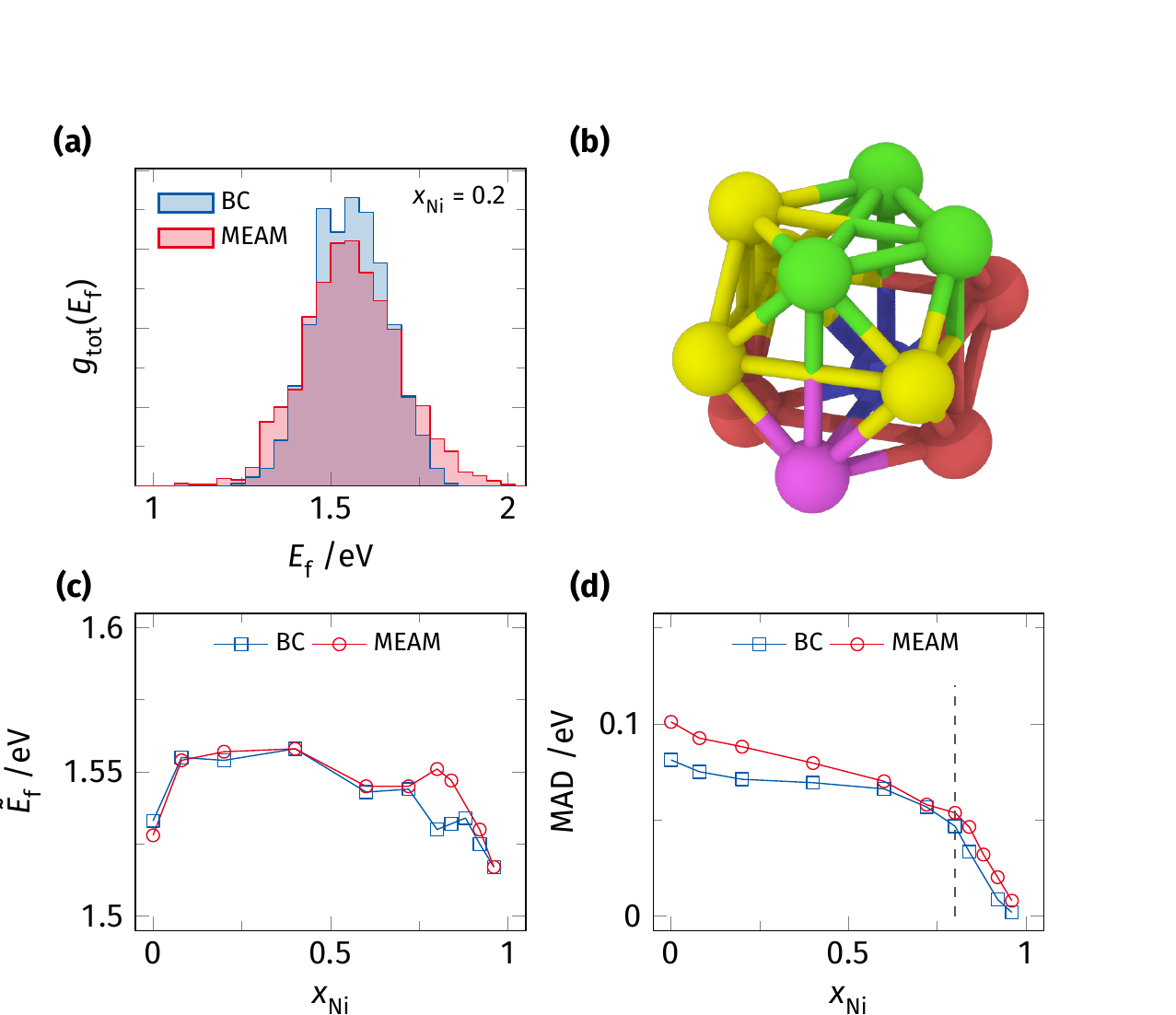}
\caption{
\textcolor{black}{
  (a) Distribution of vacancy formation energies \(g_\mathrm{tot}(E_\mathrm{f})\) calculated using the bond counting (BC) model and the full MEAM interatomic potential.
  (b) Local environment and bond topology around each vacancy that is considered in the BC model.
  (c) Median vacancy formation energy \(\widetilde{E}_\text{f}\) as function of Ni concentration \(x_\mathrm{Ni}\) obtained from the full MEAM model in comparison to the derived BC model. 
  (d) M median average deviation (MAD) \cite{Leys2013} comparing the width of the vacancy formation energy distributions between both models. The FCC site percolation threshold is indicated \cite{Gaunt1983}.
  }
\label{fig3}}
\end{figure*}

\subsection{Vacancy formation energy}

The vacancy formation energies for the different elements are calculated using \(\mu_i\) and \autoref{eq_diff_chemPot}. Following the analysis of 8000 different vacancies (1600 per removed species) for each Ni concentration, we calculate the distribution of vacancy formation energies. \autoref{fig1}(c) shows the obtained distributions \(g_i(E_\text{f})\) for the equimolar CoCrFeMnNi sample and vacancy type. The data is color coded according to the removed atom's type \(i\). 
As expected, the distributions of \(E_\text{f}\) at a given alloy composition are fairly independent of the removed species as they share almost identical median and median absolute deviation (cf.\ \autoref{fig2} (a)). 
The site specific values \(E_\mathrm{f}\), however, span a range from \SI{1}{\electronvolt} to \SI{2}{\electronvolt}. This finding is also in line with the results of Ref.~\cite{Ruban2016}, who report mean vacancy formation energies independent of the removed atom's species in a equimolar binary alloy.  \autoref{fig1} (d) shows how the distribution of \(E_\text{f}\) changes with composition for all \Nix{1-x}{x}samples.
Here, the total distribution of vacancy formation energies  \(g_\text{tot} = \sum_i^N x_i g_i\) is plotted, where \(x_i\) is the concentration and \(g_i\) is the distribution of vacancy formation energies of atom type \(i\). Note, each \(g_i\) is normalized such that \(\int g_i d E_\text{f} = 1\), while the vacancy formation energy of pure Ni \(x_\text{Ni} = 1\) is given as reference.
The transition from the concentrated (left) to the dilute (right) alloy mainly leads to a change of the peak shape, transitioning form a singular broad peak into multiple discrete peaks (\autoref{fig1} (d)).
This can be explained by a transition from a random chemical environment of the vacancy to a predominantly Ni-rich one. The highest peak in the dilute alloys corresponds to the vacancy formation energy in pure Ni (\SI{1.51}{\electronvolt}) while the different smaller peaks correspond to cases, where the vacancy is close to one or more non-Ni solutes. 

\autoref{fig2}(a\&b) show the median vacancy formation energy  \(\widetilde{E}_\text{f}\) and median average deviation (MAD) as function of alloy composition in the \Nix{1-x}{x} system  and average deviation. It can be seen that the mean vacancy formation energy is identical for all elements and remains almost constant (\(\approx \SI{1.55}{\electronvolt}\)) for all Ni concentrations. 
Also the MAD exhibits a weak dependence on the Ni content for \(x_\mathrm{Ni}<0.8\), while it converges to the single value of pure Ni for higher Ni contents in the alloy. This change in distribution corresponds to the transition from concentrated to the dilute solid solution. It occurs around the site percolation threshold of the FCC lattice, at \(x\approx\num{0.2}\), for non-Ni atoms \cite{Gaunt1983}. Once the concentration of non-Ni atoms drops below this threshold they become isolated and no \textit{network} of non-Ni atoms can form throughout the material. Therefore, these secondary atoms stop interacting as envisioned for a HEA and instead behave like dilute, isolated solutes. This fact is illustrated in 
\autoref{fig2}(c) where, once the percolation threshold is surpassed \(x_\mathrm{Ni} > 0.8\), the fraction of isolated non-Ni atoms grows rapidly.

To summarize, we have shown that in the \Nix{1-x}{x} alloy system the vacancy formation energy is  independent of the removed atom but is distributed due to different chemical environments. In dilute alloys, this distribution of formation energies features multiple discrete peaks with the highest one corresponding to the vacancy formation energy in the pure metal. The concentrated alloy, on the other hand, exhibits a single broad distribution as there are many different chemical environments leading to many different vacancy formation energies.

\subsection{Numerical Assessment of Vacancy Concentrations}

In a next step we perform grand-canonical (GC) lattice Monte-Carlo simulations and compare the simulated vacancy concentrations to the various model predictions described in Section \ref{sec_vac_meth_conc}. Since the simulations sample the configurational entropy directly and account for the energetic differences between different possible vacancy sites, the numerical result is independent of specific assumptions on how to treat configurational entropy contributions, and thus describes the relative probabilities of possible microstates in a statistically correct manner. 

\begin{figure*}[tb]
\centering
\includegraphics{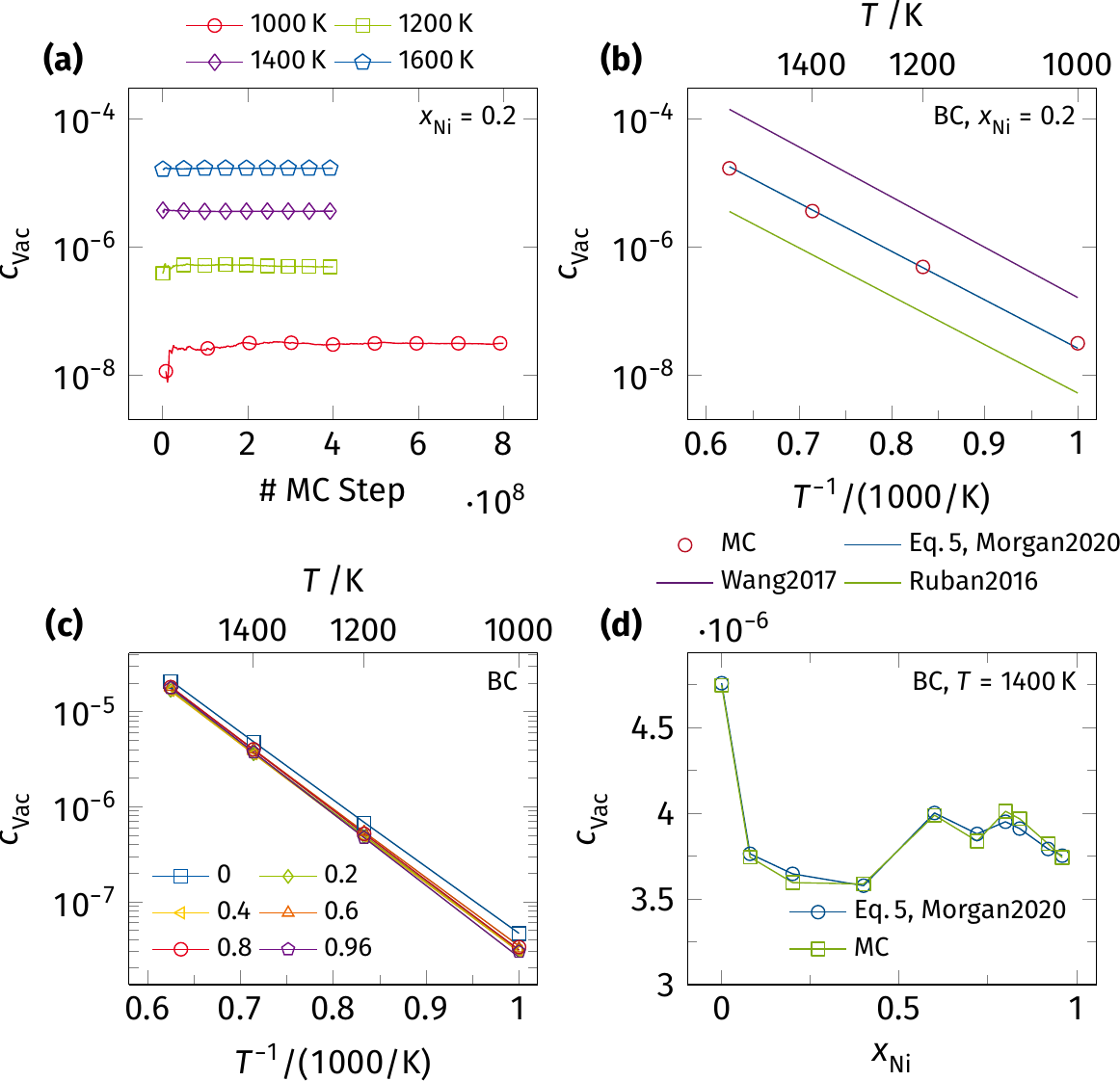}
\caption{
\textcolor{black}{
  (a) Cumulative average vacancy concentration \(c_\mathrm{Vac}\) over the number of MC steps at different temperatures for the equimolar \Nix{1-x}{x} Cantor alloy. 
  (b) Comparison of the vacancy concentrations calculated based on the equations by Morgan and Zhang (\autoref{eqVacM}) \cite{Morgan2020} Ruban \cite{Ruban2016}, and Wang et al.\ \cite{Wang2017} in comparison to our grand-canonical Monte-Carlo (MC) simulations. The resulting vacancy concentrations for the equimolar Cantor alloy (\(x_\mathrm{Ni}=0.2\)) are shown at different temperatures.
  (c) Arrhenius plot of the vacancy concentrations obtained from the BC model as function of temperature and composition of the alloy. Solid lines are calculated based on \autoref{eqVacM} while symbols mark discrete points calculated using MC\@. 
  (d) shows a vertical slice at \(T=\SI{1400}{\kelvin}\) across the different compositions highlighting that the alloy composition has a much weaker effect on vacancy concentration than temperature.
  }\label{fig4}}
\end{figure*}
\emph{Bond counting model:}
As the full MEAM interatomic potential cannot be evaluated sufficiently fast for these MC simulations, we first fit the bond counting (BC) model described in Section \ref{sec_vac_meth_conc}. 
\autoref{fig3}(a) shows the distribution of vacancy formation energies in the equimolar Cantor alloy (\(x_\text{Ni} = 0.2\)) sample as calculated with the MEAM interatomic potential and the BC model for comparison. Overall good agreement is achieved with the BC model slightly underestimating the width of the distribution. A snapshot of the chemical environment considered in the BC model featuring bonds between nearest and next-nearest neighbors in the first shell around the vacancy is shown in \autoref{fig3}(b) with vacancy the center of this shell.
\autoref{fig3}(c\&d) show the median vacancy formation energy and the median average deviation (MAD) calculated based on the full MEAM interatomic potential and the BC model.
Even though, the BC model slightly underestimates the MAD, i.e., the width of the distribution, it still approximates the real alloy system sufficiently well to allow for meaningful comparisons within the BC reference system. Especially, the important transition from concentrated to dilute solid solution is captured.

\emph{Monte Carlo simulation}:
\autoref{fig4}(a) shows the average of the vacancy concentration over MC steps for temperatures between \SI{1000}{\kelvin} and \SI{1600}{\kelvin} in the equimolar (\Nix{1-x}{x}) alloy. 
A direct comparison of the vacancy concentrations in the equimolar CoCrFeMnNi Cantor alloy calculated using the distinct thermodynamic models and the GC MC algorithm is presented in \autoref{fig4}(b). A significant difference of either a factor of \(1/5\) or \(10.9\), as predicted by the models of Refs.~\cite{Ruban2016} or \cite{Wang2017}, respectively, would be expected if the configurational entropy of the host lattice influenced the equilibrium vacancy concentration. However, the plot clearly shows the excellent agreement of the MC simulation results with the model proposed by Morgan and Zhang (\autoref{eqVacM}) \cite{Morgan2020}. This proofs that the vacancy concentration is independent of the configurational entropy of the host alloy.

Based on the previous paragraph one might wonder how different atomic configurations were sampled. To this end, we set up additional MC simulations where random swaps of the occupied lattice sites were alternated with the vacancy insertion/removal steps. This accounts for different high temperature configurations of the HEA matrix within the BC MC framework. While this approach neglects the energy of mixing in a given alloy it approximates the ideal high temperature state of the random solid solution alloys. This additional swapping of atoms, however, did not result in any change in equilibrium vacancy concentration.

\begin{figure}[tb]
\centering
\includegraphics{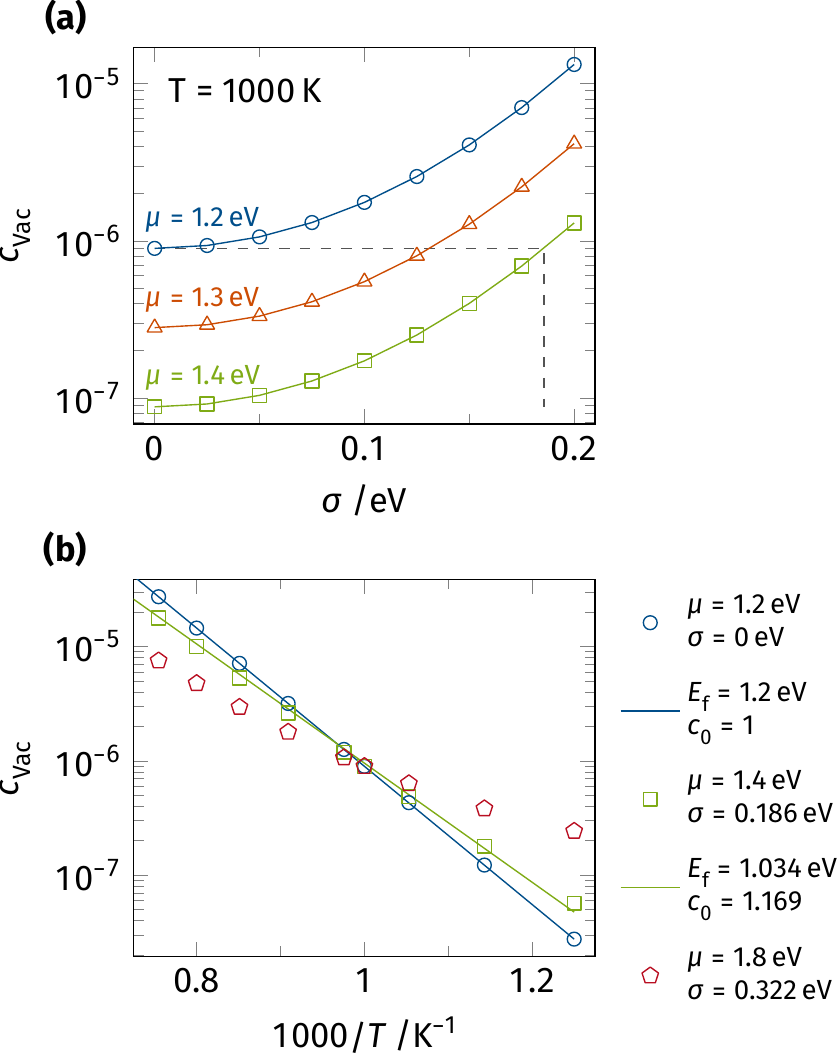}
\caption{
  (a) Vacancy concentration \(c_\text{Vac}\) as function of the vacancy formation energy distribution. This distribution is assumed to be Gaussian with different mean values \(\mu\) (\SIrange{1.2}{1.4}{\electronvolt}) and standard deviations \(\sigma\). The resulting vacancy concentration is calculated based on \autoref{eqVacM} with the entropy prefactor set to \(1\), i.e., \(S_\text{f} = 0\). Different combinations of \(\mu\) and \(\sigma\) lead to the same vacancy concentration (indicated by dashed lines).
  (b) Arrhenius plot of the vacancy concentration calculated for different combinations of \(\mu\) and \(\sigma\). The different parameter sets (circle, square, and pentagon) lead to the same vacancy concentration at \SI{1000}{\kelvin}. They do, however, diverge for other temperatures. Moreover, a wider vacancy formation energy distributions leads to a deviation from the ideal Arrhenius behavior.\label{fig5}}
\end{figure}

\autoref{fig4}(c\&d) show the resulting equilibrium vacancy concentration determined from MC simulations and \autoref{eqVacM} as function of temperature and alloy composition. The Arrhenius plot shows that the vacancy concentration changes orders of magnitude as the temperature is increased from \SI{1000}{\kelvin} to \SI{1600}{\kelvin} (c). 
A vertical slice at \(T=\SI{1400}{\kelvin}\) is given in \autoref{fig4}(d) as function of alloy composition. Here it can be seen that the equilibrium vacancy concentration varies much less with composition than with temperature.
The MC simulations confirm, that the vacancy concentration even in concentrated alloys may be calculated based on the established Maxwell-Boltzmann relation, confirming the analytical derivation by Morgan and Zhang \cite{Morgan2020}.
Moreover, our calculations show that the mean of the vacancy formation energy and thereby the vacancy concentration remains almost independent of the Ni concentration. Small differences in vacancy concentration arise from the sharpening of the vacancy formation energy distribution (see \autoref{fig5} for further details). 
The absolute change in the vacancy concentration as function of Ni content in the \Nix{1-x}{x} alloy remains small (a factor of \num{3}) and therefore hints toward the fact that differences observed in recent radiotracer experiments and associated atomistic simulations on trace diffusion are not due to significant changes in vacancy concentration \cite{Kottke2020}. 

\subsection{Role of the width of distribution}

The presented numerical analysis has confirmed that the thermodynamic model by Morgan and Zhang yields the correct equilibrium vacancy concentrations even for concentrated solid solutions. Now we investigate the implications of the width of the distribution of vacancy formation energy on the resulting vacancy concentration.

To this end, we calculate the vacancy concentration using \autoref{eqVacM} under the assumption of a Gaussian distribution with mean vacancy formation energy \(\mu\) and standard deviation \(\sigma\). Note, that for this calculation the entropy prefactor is set to \(1\), i.e., \(S_\text{f} = 0\). \autoref{fig4}(a) shows the resulting vacancy concentrations for different combinations of \(\mu\) and \(\sigma\) at a temperature of \SI{1000}{\kelvin}. It can be seen that different input parameters lead to the same equilibrium vacancy concentration. This is indicated by dashed lines showing that \(\mu = \SI{1.2}{\electronvolt}, \sigma=\SI{0}{\electronvolt}\) and \(\mu = \SI{1.4}{\electronvolt}, \sigma=\SI{0.186}{\electronvolt}\) lead to an identical concentration  \(c_\text{Vac} \approx \num{e-6}\).

Based on this observation one might be tempted to calculate an ``apparent'' vacancy formation energy which replaces the energy distribution with single value providing the same vacancy concentration.
\autoref{fig4}(b) shows in an Arrhenius plot  the vacancy concentration obtained from different combinations of \(\mu\) and \(\sigma\) (symbols).  Solid lines give the best fit to the prototypical Arrhenius equation,
\begin{equation}
  \ln c_\text{Vac} = -\frac{E_\text{f}}{k_\text{B} T}+ c_0,
\end{equation}
where \(c_0\) is the vacancy concentration at infinite temperature. Most obviously, the data for \(\mu = \SI{1.2}{\electronvolt}, \sigma=\SI{0}{\electronvolt}\) and \(\mu = \SI{1.4}{\electronvolt}, \sigma=\SI{0.186}{\electronvolt}\) intersect at \SI{1000}{\kelvin} as predicted from \autoref{fig4}(a). For lower and higher temperatures, however, the vacancy concentrations diverge. 
Comparing the two linear fits shows that the Arrhenius fit of the \(\mu = \SI{1.2}{\electronvolt}, \sigma=\SI{0}{\electronvolt}\) data set also gives an \(E_\text{f}\) of \SI{1.2}{\electronvolt}.
The \(\mu = \SI{1.4}{\electronvolt}, \sigma=\SI{0.186}{\electronvolt}\) data set, on the other hand, only gives an effective \(E_\text{f}\) of \SI{1.034}{\electronvolt} based on the fit. The individual data points also reveal a slight deviation from the fit line, i.e., the ideal Arrhenius behavior.

Lastly, to show that the deviation from the ideal Arrhenius behavior stems from the vacancy formation energy distribution width we construct a more extreme case with \(\mu = \SI{1.8}{\electronvolt}, \sigma=\SI{0.322}{\electronvolt}\), which has again a vacancy concentration of \(\approx \num{e-6}\) at \SI{1000}{\kelvin}. Comparing this new data set to the previous ones reveals a strong deviation of from the ideal linear behavior with a kink at \SI{1000}{\kelvin} where the slope changes visibly.

To conclude this section, we saw that even though there are many different combinations of mean and standard deviation leading to the same vacancy concentration at a single temperature. These different parameters give diverging vacancy concentrations at other temperatures.
Therefore, the vacancy formation energy distribution may not be approximated by a scalar effective vacancy concentration.

\section{Discussion}
Looking at available DFT calculations of vacancy formation energies in the Cantor alloy, e.g. Mizuno et al.\ \cite{Mizuno2019} or Guan et al.\ \cite{Guan2020}, shows that while it is possible to obtain mean vacancy formation energies for five component alloys using these computationally costly methods, the sample size and number samples that can be calculated is too small to converge the full vacancy formation energy distributions.
To remedy this issue we decided to perform calculations based on the Choi et al.~\cite{Choi2018} classical interatomic potential to determine \num{90000} vacancy formation energies, a number unattainable to DFT calculations.
Now, one might ask how well this interatomic potential performs compared to the available DFT data. A graphical comparison is given in \textcolor{blue}{Figure~S1}. Here we can conclude, that even though, the two references do not agree on the elemental hierarchy, their vacancy formation energy is on the order of \SI{2}{\electronvolt} which is substantially higher than the \SI{1.55}{\electronvolt} determined from the empirical potential, as shown in \autoref{fig2}(a).
While this difference in mean vacancy formation energy changes the absolute vacancy concentration at a given temperature it does not meaningfully influence the conclusions of this work. 
The concentration dependence of the vacancy formation energy distribution is related to number atoms of a given species surrounding a vacant site. At least in an ideally random alloy this is a purely statistical effect (cf.\ \autoref{fig2}(c) \& \textcolor{blue}{Figure~S2}) and it is therefore independent of the interatomic potential. Similarly, the transition from concentrated to dilute solid solution is marked by the percolation of ``solute'', i.e., non-Ni atoms within the \Nix{1-x}{x} alloy system. This percolation threshold of the FCC lattice is a well established mathematical quantity for the FCC lattice and does not relate to the properties of the atoms occupying the lattice. 
Lastly, the comparison of the different thermodynamic models to the GC MC simulations will give ``wrong'' absolute vacancy concentrations, at least in comparison the experiments on the Cantor alloy, however, as both thermodynamic models and MC simulations are within the same context of the BC model the conclusions hold, independent of any errors in the interatomic potential.

Another difference between our calculations and experimentally obtained vacancy concentrations for this system lies in the treatment of vibrational entropies. All our calculations are based on the assumption of \(S_\text{f} = 0\). 
A finite but concentration independent value of \(S_\text{f}\) would again shift the absolute vacancy concentrations (cf.\ \autoref{eqVacM}) without changing the conclusion of \autoref{fig3}(f) where we showed that the vacancy concentration is almost independent of alloy composition within the \Nix{1-x}{x} system.
This is not an unreasonable assumption given that the lattice vibrational entropy of vacancy formation is almost constant for many pure metals \cite{Burton1972} and concentration independent for binary alloys \cite{Mosig1992,Hehenkamp2001}. 
Similarly, Aziziha and Akbarshahi \cite{Aziziha2020} have calculated the vibrational entropy for defect free CoFeNi, CoCrFeNi, and CoCrFeMnNi and reported values between \(8\,k_\text{B}\) and \(8.4\,k_\text{B}\), indicating only a weak concentration dependence.
Magnetic entropy contributions have the smallest effect on the vacancy formation energy and are only relevant above \SI{80}{\percent} of the melting temperature \cite{Gong2018}.

There is, however, another open problem involved with the step from  distribution of formation energies to vacancy concentration, which is still unresolved. It has to do with the chemical inhomogeneity of the concentrated solid solution.
The vacancy formation energies shown in \autoref{fig1} are determined from ideally random atomic configurations. The shape of the vacancy formation energy distribution might change, however, if there is some degree of short range ordering in a given alloy. For many HEAs it is still unclear whether they form ideally random solid solutions based on established processing routes or whether the atoms have sufficient time to stabilize a finite degree of ordering. So this effect cannot be excluded for comparison with the experiments.
The chemical disorder also leads to a secondary issue. The conventional derivation of the equilibrium vacancy concentration is based on the Boltzmann entropy of mixing for host atoms and vacancies. However, this equation only holds, if all microstates, i.e., vacancy sites, are energetically equivalent. This assumption does not hold for alloys, as shown in \autoref{fig1}. Therefore, the Gibbs entropy formulation needs to be used instead. Here each microstate is weighted by its probability which in turn depends on its energy. However, to our knowledge such derivation has not been attempted yet.

An important conclusion of our work, for the comparison of calculated and measured vacancy formation energies, is that they are fundamentally not comparable. From the analysis in \autoref{fig4}(b) it becomes clear that the vacancy formation energy determined from an Arrhenius plot of experimentally measured vacancy concentrations, obtained from positron annihilation spectroscopy or dilatation measurements, are always effective quantities encompassing the full vacancy formation energy distribution. 
Moreover, this effective vacancy formation energy differs from the atomistic vacancy formation energies determined from calculations. In the example shown here the atomic vacancy formation energies are \SI{1.4 +- 0.186}{\electronvolt} while the effective vacancy formation energies determined from the corresponding, fictitious, experiment is equal to \SI{1.034}{\electronvolt}.

\section{Conclusion}
We found that the vacancy formation energy in alloys is not a singular value but follows a distribution. 
In the \Nix{1-x}{x} system, the median of these distributions is almost constant with composition while the width and shape changes substantially as the alloy transitions from concentrated to dilute solid solution.
Moreover, in line with previous findings and general thermodynamic rules, the vacancy formation energy is independent of the removed atom's species.

Subsequent MC simulations reveal perfect agreement of the vacancy concentration and the thermodynamic model by Morgan and Zhang \cite{Morgan2020}.
This confirmation allows us to conclude that within the \Nix{1-x}{x} alloys, the vacancy concentration changes only weakly with the alloy composition.
Using this thermodynamic model we are able to show that, even though, vacancy formation energies with different means and standard deviations can lead to the same vacancy concentration at a given temperature they are non-equivalent at all other temperatures.
 This means that a system with a vacancy formation energy distribution, the experimentally measured effective vacancy formation energy is not equivalent to the vacancy formation energy calculated using atomistic methods.

\section*{Acknowledgement}
The authors would like to acknowledge financial support by the Deutsche Forschungsgemeinschaft (DFG) under grant Nos.\ STU 611/2-1 \& Al 578/25-2 as part of the SPP 2006.
Calculations for this research were conducted on the Lichtenberg high performance computer of the TU Darmstadt.
The authors gratefully acknowledge the Gauss Centre for Supercomputing e.V.\ for funding this project by providing computing time on the GCS Supercomputer SuperMUC-NG at Leibniz Supercomputing Centre. The authors acknowledge helpful discussions with A.J.\ Klomp and M.\ Sadowski.
\bibliographystyle{elsarticle-num}
\bibliography{literature.bib}

\end{document}

\title{Supplementary Material\\
Thermodynamics of vacancies in concentrated solid solutions:\\
From dilute Ni-alloys to the Cantor system}

\author[mm]{Daniel Utt \corref{cor1}}
\ead{utt@mm.tu-darmstadt.de}
\cortext[cor1]{Corresponding author:}
\author[mm]{Alexander Stukowski}
\author[mm]{Karsten Albe}
\address[mm]{Technische Universit\"at Darmstadt, Otto-Berndt-Str.\ 3, 64287 Darmstadt, Germany}

\maketitle

\begin{figure}[H]
  \centering
  \includegraphics{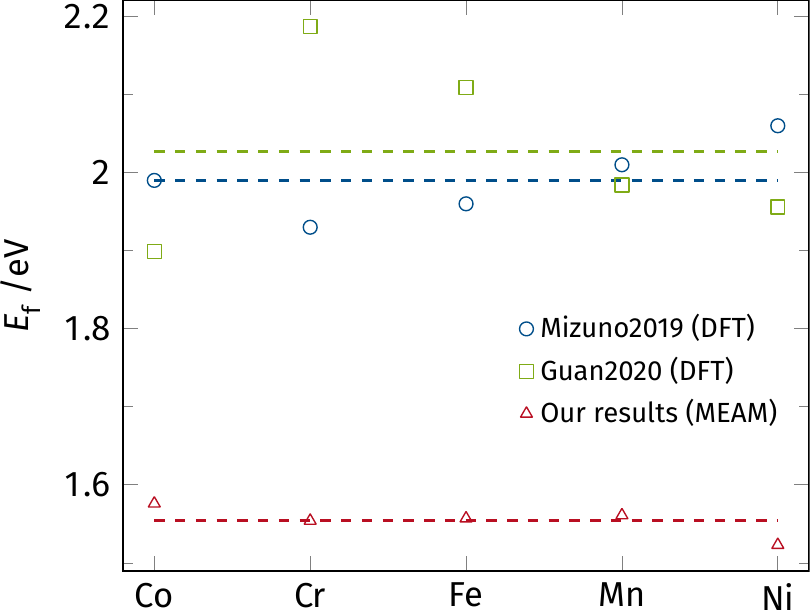}
  \caption{
  \textcolor{black}{
    Comparison of the vacancy formation energy we calculated from the MEAM interatomic potential \cite{Choi2018} in comparison to DFT calculations by Mizuno et al.\ and Guan et al.\ \cite{Mizuno2019,Guan2020}. Dashed liens give the respective mean. Note how our results show a much better convergence towards a single mean vacancy formation energy, independent of the removed atom's species.
  \label{fig2-5_supp}}}
\end{figure}

\begin{figure}[H]
\centering
\includegraphics{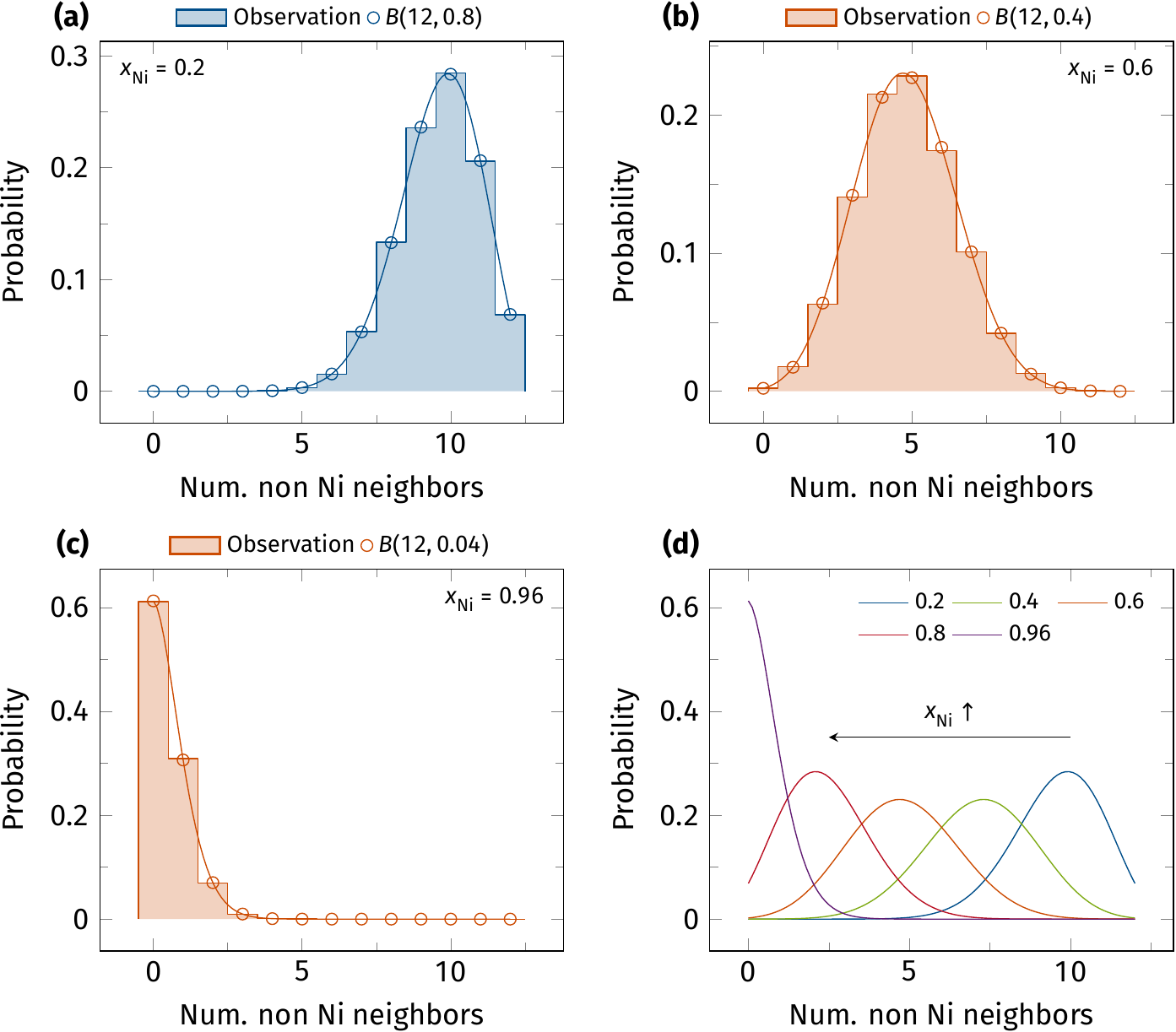}
\caption{
  Binomial distribution \(B(n,p) = \binom{n}{k} p^k (1-p)^{n-k}\), where \(n\) is the number of neighbors (\(n = 12\) for the FCC crystal structure), k is the number of non Ni neighbors and \(p\) is the concentration of non Ni species (\(p = 1-x_\mathrm{Ni}\)).
  (a-c) Comparison between the number of non Ni atoms surrounding a vacancy site observed in our samples compared to the expectations for a random sample from the binomial distribution for selected \(x_\mathrm{Ni}\). Perfect agreement between observation and analytical description can be seen.
  (d) Probability for a given number of non Ni neighbors around a vacancy summarized for many Ni concentrations.
  \label{fig1_supp}}
\end{figure}

\bibliographystyle{elsarticle-num}
\bibliography{literature.bib}